\documentstyle[prl,aps,amsfonts,amssymb,twocolumn,epsfig]{revtex}

\newcommand{\nm}{\rm}
\newcommand{\so}{\rm }

\newcommand{\JETPL}{JETP Lett.}
\newcommand{\PZETF}{Pis'ma Zh.\ Eksp.\ Teor.\ Fiz.}

\newcommand{\EL}{Phys.\ Rev.\ Lett.}

\newcommand{\JPCM}{J.\ Phys.: Cond.\ Matter}

\newcommand{\PR}{Phys.\ Rev.}
\newcommand{\PRB}{Phys.\ Rev.\ B}

\newcommand{\PRL}{Phys.\ Rev.\ Lett.}

\newcommand{\RMP}{Rev.\ Mod.\ Phys.}

\newcommand{\ZPB}{Z.\ Phys.\ B}

\begin{document}
\title{
Anderson impurity in pseudo-gap Fermi systems}
\author{R.~Bulla$^1$, Th.~Pruschke$^2$ and A.~C.~Hewson$^1$}
\address{
$^1$Department of Mathematics, Imperial College,\\
180 Queen's Gate, London SW7 2BZ, United Kingdom\\
$^2$Institut f\"ur Theoretische Physik der Universit\"at,\\
93040 Regensburg, Germany
}
\draft

\twocolumn[\hsize\textwidth\columnwidth\hsize\csname
@twocolumnfalse\endcsname %

\date{\today}

\maketitle
\begin{abstract}
We use the numerical renormalization group method to study an
Anderson impurity in a conduction band with the
density of states varying as $\rho(\omega) \propto |\omega|^r$ 
with $r>0$.
We find two different fixed points: a local-moment fixed point
with the impurity effectively decoupled from the band and a strong-coupling
fixed point with a partially screened impurity spin. 
The specific heat and the spin-susceptibility show powerlaw behaviour
with different exponents in strong-coupling and local-moment regime.
We also calculate the impurity spectral function which 
diverges (vanishes) with  $|\omega|^{-r}$ ($|\omega|^r$)
in the strong-coupling (local moment) regime.

\end{abstract}

\vskip1.0pc]

\section{Introduction}

The behaviour of a magnetic impurities in metals is one of the best studied
problem in condensed matter theory \cite{Hew93}.
In most cases it is a very good approximation to replace the conduction
density of states by a constant as small variations of the density of states
do not lead to a qualitative change of the physical properties 
(like the complete screening 
of the inpurity spin by the conduction electrons).

The question, whether these physical properties are different when
the impurity is coupled to a Fermi system with a power-law
density of states $\rho(\omega) \propto |\omega|^r$
near the Fermi-level was first discussed by Withoff
and Fradkin \cite{Wit90}.

A number of systems are expected to show this pseudogap density of states.
Among these are certain heavy-fermion superconductors 
\cite{Sig91} where the exponent $r$ can take the values $r=1$ or $r=2$
depending on the symmetry of the gap function.
Other candidates are semiconductors whose valence
and conduction bands touch at the Fermi level \cite{Vol85}.
In quasi one-dimensional metals,
which can be viewed as realizations of the Luttinger
model, the exponent $r$ is a function of the
Coulomb interaction\cite{Dar93a} and can take
values between $r\! <\!< \!1$ and $\!r>\!1$.

Recently, the numerical renormalization group method (NRG) 
\cite{Wil75,Kri80}
has been applied by Chen and Jayaprakash \cite{Che95} (referred to as CY)
and Ingersent \cite{Ing96} 
to the model of an impurity spin coupled to a conduction band
with a powerlaw density of states. In principal, this Kondo
model can be related to a corresponding Anderson model
in the limit of $J\to 0$ via a standard Schrieffer-Wolff transformation
\cite{Sch66}.
There is, however, a transition between a strong-coupling (SC)
fixed point and a local-moment (LM) fixed point for
the Kondo model at {\it finite}
$J$ so that it is a priori not clear whether the behaviour
at this transition will be the same in the Anderson
version of the model.

The results of CY and Ingersent can be summarized as follows.
For any $J>J_{\rm c}$ the system approaches some kind of SC fixed point
with the difference to the standard Kondo model ($r=0$) that
the impurity spin is not completely screened (a residual magnetic
moment of $r/8$ always remains in the zero-temperature limit).
This can be qualitatively understood from the gradually decreasing
density of states of the conduction electrons at the
Fermi level which are responsible for the screening.

The thermodynamic quantities show non-Fermi liquid behaviour in the
SC regime
\begin{eqnarray}
    \gamma(T) &=& \frac{C(T)}{T} \propto  T^{-r},    \\      
   \chi_S(T) &=& \frac{r}{8} T^{-1} + a T^{-r}+ b T^{-2r}, 
\end{eqnarray}
(with $a,b =$ const.).
The critical line $J_{\rm c}(r)$ starts linearly for small $r$ but diverges
at $r=\frac{1}{2}$. In addition, Ingersent has shown that this divergence
only holds
in the particle-hole symmetric case and that a finite $J_{\rm c}$
is restored away from this symmetry. 
(This reduction of $J_{\rm c}$ has implications for the
observability of the crossover in experimental situations.)

For any $J<J_{\rm c}$, the system approaches the LM fixed point
where the impurity is effectively decoupled from the
conduction band and a residual magnetic moment of $1/4$ remains.
The thermodynamics in this regime have not yet been investigated.

In this paper, we want to study the behaviour of an Anderson impurity
in a pseudo-gap fermion system where we restrict ourselves to the 
symmetric case.
In Sec.\ II, we want to describe our approach to the generalization of 
the NRG with a non-constant density of states, and outline
the differences to that of CY and Ingersent.
The resulting formula for the hopping matrix elements
of the semi-infinite chain for {\it all $n$} is given in Sec.\ III.
The numerical results for static properties and the spectral function
are discussed in Sec.\ IV and V, respectively.

\section{Generalization of NRG to nonconstant density of states}
The Hamiltonian we want to study in this paper is the conventional
single-impurity Anderson model
\begin{eqnarray}
  H &=&   \sum_{\sigma} \varepsilon_{\rm f} f^\dagger_{-1 \sigma}
                             f_{-1 \sigma}
                 + U  f^\dagger_{-1 \uparrow} f_{-1 \uparrow}
                       f^\dagger_{-1\downarrow} f_{-1\downarrow}
                \nonumber \\
           &+& \sum_{k \sigma} \varepsilon_k c^\dagger_{k\sigma} c_{k\sigma}
            +  \sum_{k \sigma} V(\varepsilon_k)
           \Big( f^\dagger_{-1 \sigma} c_{k \sigma}
               +   c^\dagger_{k\sigma} f_{-1\sigma} \Big). 
    \label{eq:siam}
\end{eqnarray}
In the model (\ref{eq:siam}), $c_{k\sigma}^{(\dagger)}$ denote standard
annihilation
(creation) operators for band states with 
spin $\sigma$ and energy $\varepsilon_k$,
$f_{-1,\sigma}^{(\dagger)}$
those for impurity states with spin $\sigma$ and energy $\varepsilon_{\rm f}$. The
Coulomb interaction for two electrons at the impurity site is given by $U$ and
both subsystems are coupled via an energy dependend hybridization
$V(\varepsilon_k)$ \cite{comment}.

In the following we show that the Hamiltonian (\ref{eq:siam}) is
equivalent to a form which is more convenient for the derivation of 
the NRG equations 
\begin{eqnarray}
  H &=&   \sum_{\sigma} \varepsilon_{\rm f} f^\dagger_{-1\sigma}
                             f_{-1\sigma}
                 + U  f^\dagger_{-1 \uparrow} f_{-1 \uparrow}
                       f^\dagger_{-1 \downarrow} f_{-1 \downarrow}
                \nonumber \\
           &+& \sum_{ \sigma}\int_{-1}^1 {\rm d} \varepsilon \, g(\varepsilon)
                  a^\dagger_{\varepsilon \sigma} a_{\varepsilon
             \sigma}\nonumber \\
            &+&  \sum_{ \sigma} \int_{-1}^1 {\rm d} \varepsilon \,
                   h(\varepsilon) \Big( f^\dagger_{-1 \sigma}
                   a_{\varepsilon \sigma}  +
                     a^\dagger_{\varepsilon \sigma} f_{-1\sigma} \Big),
    \label{eq:siam_cont}
\end{eqnarray}
where we introduced a one-dimensional energy
representation for the conduction band with band cut-offs at $\pm 1$,
dispersion
$g(\varepsilon)$ and hybridization $h(\varepsilon)$. The band operators fulfil the
standard fermionic commutation rules $\left[a_{\varepsilon\sigma}^{\dagger},
a_{\varepsilon'\sigma'}\right]=\delta(\varepsilon-\varepsilon')
\delta_{\sigma\sigma'}$.

To establish the equivalence of the Hamiltonians (\ref{eq:siam}) and
(\ref{eq:siam_cont})
we prove that for a specific choice of $g(\varepsilon)$ and $h(\varepsilon)$ they lead
to the same effective action for the impurity degree of freedom. 
This effective
action is obtained by integrating over the conduction electron
degrees of freedom. For the Hamiltonian (\ref{eq:siam}) one gets
\begin{eqnarray}
S_{\rm eff}(\psi,\psi^\dagger)
  &=& S_{\rm f}(\psi,\psi^\dagger)  \nonumber \\
 & & \!\!\!\!\!\!\!\!\!\!\!\! \left( {{\beta } \over N} \right)^2
\sum_{\sigma  n  m}  \psi^\dagger_{\sigma n+1}
\psi_{\sigma m-1}   \sum_k V(\varepsilon_k)^2
G^c_{n m}(k),      \label{eq:Seffk}
\end{eqnarray}
(see for example \cite{Bul94}). $n$ and $m$ count the steps on the imaginary
time axis $[0,\beta]$ with $N$ the number of steps. 
$\psi$ and $\psi^\dagger$ are Grassmann numbers corresponding to the
impurity operators. $S_{\rm f}$ describes the unhybridized impurity.
The $G^c_{n m}(k)$ are Green functions for the free conduction electron
system.

The action corresponding to the Hamiltonian (\ref{eq:siam_cont})
can be written as
\begin{eqnarray}
S (\psi,\psi^\dagger,\chi,\chi^\dagger)
  &=& S_{\rm f}(\psi,\psi^\dagger) \nonumber \\ & & 
        \hspace{-2cm} +
      \sum_{\sigma n} \int_{-1}^1 {\rm d} \varepsilon \,
         \chi_{\varepsilon\sigma n}^\dagger  \Big(
         \big( 1 -  \frac{\beta}{N} g(\varepsilon) \big)
         \chi_{\varepsilon\sigma n-1} - \chi_{\varepsilon\sigma n} \Big)
     \nonumber \\
  & &  \hspace{-2cm} -
        \frac{\beta}{N} \sum_{\sigma n} \int_{-1}^1 {\rm d} \varepsilon \,
         h(\varepsilon) \Big[
  \chi_{\varepsilon\sigma n}^\dagger  \psi_{\sigma n-1} +
  \psi^\dagger_{\sigma n} \chi_{\varepsilon \sigma n-1} \Big].
\end{eqnarray}
$\chi_{\varepsilon\sigma n}^\dagger $ and 
$\chi_{\varepsilon \sigma n}$ are Grassmann numbers corresponding to the
conduction electron operators 
$a^\dagger_{\varepsilon \sigma}$ and $a_{\varepsilon \sigma}$ .
Integrating over the conduction electron degrees of freedom leads to
\begin{eqnarray}
S_{\rm eff}(\psi,\psi^\dagger)
  &=& S_{\rm f}(\psi,\psi^\dagger) \nonumber \\
& & \hspace{-1.5cm}
       +\left( {{\beta} \over N} \right)^2
\sum_{\sigma  n  m}  \psi^\dagger_{\sigma n+1}
\psi_{\sigma m-1} \int_{-1}^1 {\rm d} \varepsilon \,
         h(\varepsilon)^2  G^c_{n m}(g(\varepsilon)). \nonumber \\
    \label{eq:Seffeps} 
\end{eqnarray}
To compare the effective actions (\ref{eq:Seffk}) and (\ref{eq:Seffeps})
the sum over $k$ in (\ref{eq:Seffk}) has to be transformed 
to the energy integral
\begin{equation}
\sum_k V(\varepsilon_k)^2
G^c_{n m}(k)    = \int_{-1}^1 {\rm d} \varepsilon \,
         V(\varepsilon)^2 \rho(\varepsilon)  G^c_{n m}(\varepsilon).
\end{equation}
This also defines the density of states for the free conduction electrons
$\rho(\varepsilon)$.
The equivalence of the effective actions  (\ref{eq:Seffk})  and
(\ref{eq:Seffeps})
leads to the condition
\begin{equation}
\int_{-1}^1 {\rm d} g \frac{\partial\varepsilon(g)}{\partial g}
          h(\varepsilon(g))^2
       G^c_{n m}(g)  \equiv
   \int_{-1}^1 {\rm d} \varepsilon \,
         V(\varepsilon)^2 \rho(\varepsilon)  G^c_{n m}(\varepsilon) 
     .
\end{equation}
This can only be fulfilled for
\begin{equation}
                \frac{\partial \varepsilon (x)}{\partial x}
                   h(\varepsilon(x))^2 = V(x)^2 \rho(x), \label{eq:diffeq}
\end{equation}   
(with $\varepsilon(x)$ the inverse of  $g(\varepsilon)$).
For a given $\Delta(x) \equiv \pi V(x)^2 \rho(x)$ there are obviously many ways of
dividing the energy dependence between $\varepsilon(x)$ and the dispersion
$h(\varepsilon(x))$.
One possibility is to choose
\begin{equation}
       g(\varepsilon) = \varepsilon \ \ \ \ {\rm and} \ \ \ \ \ 
    h(\varepsilon)^2 = \frac{1}{\pi}\Delta(\varepsilon). \label{eq:pos1}
 \end{equation}   
For $\Delta(\varepsilon) = \Delta$ eq.\ (\ref{eq:pos1}) corresponds to the
standard case (see eq.\ (2.4) in \cite{Kri80}).
It might also be convenient to set $h(\varepsilon)=h$.
Together with the condition $\varepsilon(-1)=-1$ and $\varepsilon(1)=1$
this leads to
 \begin{displaymath}
 \varepsilon(g) = -1 + \frac{1}{\pi h^2}\int_{-1}^{g} {\rm d} x \Delta(x)
  \ \ \ \ {\rm and} 
\end{displaymath}
\begin{equation}
   h^2=\frac{1}{2\pi}\int_{-1}^{1} {\rm d} \varepsilon \Delta(\varepsilon) 
   . 
  \label{eq:pos2}
\end{equation}
This equations also reduce to $\varepsilon(g) = g$ and $h^2 = \frac{1}{\pi}\Delta$
for a constant $\Delta(\varepsilon) = \Delta$.
Equations (\ref{eq:pos1}) and (\ref{eq:pos2}) have already been derived by
CY \cite{Che95a}. In a subsequent publication \cite{Che95}
these authors use eq.\ (\ref{eq:pos2}) for the mapping of the Kondo model
on a semi-infinite chain (see Appendix A for a discussion of the resulting
hopping matrix elements).

The first possibility eq.\ (\ref{eq:pos1}) has a conceptual disadvantage
arising from the logarithmic discretization of the conduction band.
Within each interval $[x_{n+1},x_n]$ and $[-x_{n},-x_{n+1}]$, with $x_n=\Lambda^{-n}$,
the conduction electron operators are expressed in terms of a Fourier
expansion. As long as $h(\varepsilon)^2$ is constant in each interval, the impurity
couples only to the average component ($p\!=\!0$) of the conduction electrons.
Therefore it is reasonable to neglect all the $p\!\ne\!0$-states (this becomes exact in
the limit $\Lambda\! \to\! 1$). This line of reasoning obviously does not
hold for eq.\ (\ref{eq:pos1}).

On the other hand, the energy dependence of $\Delta(\varepsilon)$ can be 
taken into account in the hybridization by defining $h(\varepsilon)^2$ 
as the mean value 
\begin{equation}
    {h^{\pm}_n}^2 = \frac{1}{d_n} \int^{\pm} {\rm d} \varepsilon
  \frac{1}{\pi}\Delta(\varepsilon), \label{eq:h_mean}
\end{equation}
\begin{equation}
  \int^{+} {\rm d} \varepsilon \equiv
    \int_{x_{n+1}}^{x_n} {\rm d} \varepsilon, \quad
  \int^{-} {\rm d} \varepsilon \equiv
    \int_{-x_n}^{-x_{n+1}} {\rm d} \varepsilon,
\end{equation}
(with $d_n = x_n-x_{n+1}$) in each interval of the logarithmic
discretization.
This is so far not an approximation as the remaining energy dependence 
will be incorporated in the dispersion.
The advantage of an energy dependent hybridization as in eq.\ (\ref{eq:h_mean}) is that the
resulting dispersion has the form $g(\pm x_n) = \pm x_n$ for all $n$,
i.e.\ at all points $x_n$ of the logarithmic discretization.
This "linear" form
(for intermediate values $g(\varepsilon)\! =\! \varepsilon$ is not fulfilled) leads to
a scaling behaviour of the hopping matrix elements (see eq.\ (\ref{eq:H_with_t_n}))
of the form $t_n \propto \Lambda^{-n/2}$, slightly modified due to the structure
of $\Delta(\varepsilon)$. The representation eq.\ (\ref{eq:pos2}), however, leads to
a scaling with an effective $\Lambda_{\rm eff}$ not equal to $\Lambda$ which might
even depend on the number of iterations thus making the analyses (of the fixed points,
the relevant energy scale, etc.) more difficult. 

For these reasons, we take the representation eq.\ (\ref{eq:h_mean}) in the following.
This gives for the hybridization part of the discretized Hamiltonian
\begin{equation}
   H_{\rm hyb} = \sqrt{\frac{\xi_0}{\pi}} \left[
            f^\dagger_{-1\sigma}f_{0\sigma} + 
            f^\dagger_{0\sigma}f_{-1\sigma}  \right],
\end{equation}
with
\begin{equation}
    f_{0\sigma} = \frac{1}{\sqrt{\xi_0}} \sum_n \left[
          \gamma_n^+  a_{n\sigma}   + \gamma_n^- b_{n\sigma} \right],
\end{equation}
\begin{equation}
   \xi_0 = \sum_n \left(  (\gamma_n^+)^2 +(\gamma_n^-)^2 \right) 
       = \int_{-1}^1 {\rm d} \varepsilon \Delta(\varepsilon), 
\end{equation} 
\begin{equation}
           (\gamma_n^\pm)^2 = \int^{\pm}  {\rm d} \varepsilon \,
                   \Delta(\varepsilon).
\end{equation}
The discrete conduction electron operators $a_{n\sigma}$ ($b_{n\sigma}$)
for positive (negative) $\varepsilon$ correspond to those introduced in
\cite{Wil75,Kri80}.
According to the differential equation (\ref{eq:diffeq}) we should now have to solve for 
$\varepsilon(x)$ and invert $\varepsilon(x)$ to obtain the dispersion 
$x(\varepsilon) \equiv g(\varepsilon)$.
This is actually not necessary because the single-particle energies
in the conduction electron part of the
discretized Hamiltonian 
\begin{equation}
  H_{\rm c} = \sum_{n\sigma} \left[ \xi_n^+ a^\dagger_{n\sigma} a_{n\sigma}
                    + \xi_n^- b^\dagger_{n\sigma} b_{n\sigma} \right]
\end{equation}
only depend on the integral over $g(\varepsilon)$ 
\begin{equation}
  \xi_n^\pm = \frac{1}{d_n} \int^{\pm}  {\rm d} \varepsilon
           \, g(\varepsilon).
\end{equation}
It can be shown that the discrete energies $\xi_n^\pm$ are
given by
\begin{equation}
  \xi_n^\pm = \frac{\int^\pm {\rm d} \varepsilon \Delta(\varepsilon) \varepsilon}{
                   \int^\pm {\rm d} \varepsilon \Delta(\varepsilon) }
            .
\end{equation}
This equation, together with the form of the hybridization part has already
been used by Sakai et al.\ \cite{Sak94}, although no derivation was given in
their article.

The discretized Hamiltonian for the single-impurity Anderson model
 now takes the form
\begin{eqnarray}
  H &=&   \sum_{\sigma} \varepsilon_{\rm f} f^\dagger_{-1\sigma}
                             f_{-1\sigma}
                 + U  f^\dagger_{-1 \uparrow} f_{-1 \uparrow}
                       f^\dagger_{-1 \downarrow} f_{-1 \downarrow}
                \nonumber \\
           &+& \sum_{n\sigma} \left[
                  \xi_n^+
                 a^\dagger_{n\sigma} a_{n\sigma} +
                   \xi_n^-
                 b^\dagger_{n\sigma} b_{n\sigma} \right] \nonumber \\
         &+& \sqrt{\frac{\xi_0}{\pi}} \left[
            f^\dagger_{-1\sigma}f_{0\sigma} + 
            f^\dagger_{0\sigma}f_{-1\sigma}  \right].
        \label{eq:Hdisc}
\end{eqnarray}

\section{Pseudogap density of states --- Mapping on semi-infinite chain}
We now consider  a $\Delta(\omega)$ of the form
\begin{equation}
    \Delta(\omega)=\Delta_0 |\omega|^r, \ \ \ \ -1\le \omega \le 1
     .
\end{equation}   
The discrete energies $\xi_n^\pm$ of the conduction electrons
and the hybridization matrix elements $\gamma_n^\pm$
between impurity and the  conduction electrons
take the form
\begin{equation}
     \xi_n^+ = - \xi_n^- = \frac{r+1}{r+2} 
     \frac{1 -\Lambda^{-(r+2)}}{1-\Lambda^{-(r+1)}}
            \Lambda^{-n} \label{eq:xi}
\end{equation}   
and
\begin{equation}
    \left(  \gamma_n^+ \right)^2 =
   \left(  \gamma_n^- \right)^2 = \frac{\Delta_0 }{r+1} \Lambda^{-n(r+1)}
       \left(  1 - \Lambda^{-(r+1)} \right).\label{eq:gamma}
\end{equation}   
The mapping of the discretized Hamiltonian (\ref{eq:Hdisc})
onto the semi-infinite chain form
\begin{eqnarray}
  H &=&   \sum_{\sigma} \varepsilon_{\rm f} f^\dagger_{-1\sigma}
                             f_{-1\sigma}
                 + U  f^\dagger_{-1 \uparrow} f_{-1 \uparrow}
                       f^\dagger_{-1 \downarrow} f_{-1 \downarrow}
                \nonumber \\
           &+& \sum_{\sigma n=0}^\infty t_n \left[
                f^\dagger_{n\sigma}f_{n+1\sigma} +
                f^\dagger_{n+1\sigma}f_{n\sigma}
           \right] \label{eq:Hsemiinf} \\
         &+& \sqrt{\frac{\xi_0}{\pi}} \left[
            f^\dagger_{-1\sigma}f_{0\sigma} + 
            f^\dagger_{0\sigma}f_{-1\sigma}  \right], \label{eq:H_with_t_n}
\end{eqnarray}
($\xi_0=\frac{2\Delta_0 }{r+1}$) is described in \cite{Wil75} and \cite{Kri80}.
The only difference appearing here is the $r$-dependence of the
$\xi_n^\pm$ and $\gamma_n^\pm$.
(Note that in the non-symmetric case additional terms of the form
$\varepsilon_n f^\dagger_{n\sigma}f_{n\sigma}$ are generated.)
For the hopping matrix elements $t_n$ we find the following expressions.
\begin{eqnarray}
    t_n &=&  \Lambda^{-n/2} \,\frac{r+1}{r+2} \, 
    \frac{1-\Lambda^{-(r+2)}}{1-\Lambda^{-(r+1)}} 
      \left[ 1 - \Lambda^{-(n+r+1)} \right]\nonumber \\
    &\times& 
       \left[ 1 - \Lambda^{-(2n+r+1)} \right]^{-1/2}
       \left[ 1 - \Lambda^{-(2n+r+3)} \right]^{-1/2}   \label{eq:tneven}
\end{eqnarray}   
  for even $n$ and
\begin{eqnarray}
    t_n &=&
  \Lambda^{-(n+r)/2} \,\frac{r+1}{r+2} \, 
      \frac{1-\Lambda^{-(r+2)}}{1-\Lambda^{-(r+1)}} 
      \left[ 1 - \Lambda^{-(n+1)} \right]\nonumber \\
   &\times& 
       \left[ 1 - \Lambda^{-(2n+r+1)} \right]^{-1/2}
       \left[ 1 - \Lambda^{-(2n+r+3)} \right]^{-1/2} \label{eq:tnodd}
\end{eqnarray}   
for odd $n$. 
The equations (\ref{eq:tneven}) and (\ref{eq:tnodd}) have been verified
numerically and by analytical calculation of $t_0$ and $t_1$.
In the limit $n\to \infty$ (\ref{eq:tneven}) and (\ref{eq:tnodd})
reduce to
\begin{equation}
      t_n \stackrel{n\to\infty}{\longrightarrow}
        \frac{r+1}{r+2} \,
        \frac{1-\Lambda^{-(r+2)}}{1-\Lambda^{-(r+1)}}
        \Lambda^{-n/2} \left\{
           \begin{array}{lcl}
           1 &:& n\ {\rm even}\\
           \Lambda^{-r/2} &:& n\ {\rm odd}
             \end{array}
            \right.   .  \label{eq:tnred}
\end{equation}   
This limit of the hopping matrix elements has also been found
by Ingersent \cite{Ing96} although the formula for {\it all} $n$ is
not given in his paper.
The result obtained by CY is discussed in the appendix.

An analytical form of the $t_n$ for all $n\ge0$ can only be given when the powerlaw
$\Delta(\omega)=\Delta_0 |\omega|^r$ extends to the band edges. In any
experimental realization, however, we expect this powerlaw only to
be valid near the Fermi level. On the other hand, numerical studies show that
any deviation from the form (23) close to the band edges merely affects the
first coefficients, while the asymptotic behaviour again depends on $r$ only
and is given by (30). Thus the qualitative behaviour near the possible
low temperature fixed points is not affected by the exact 
form of $\Delta(\omega)$ away from the Fermi level.

\section{Results for static properties}
The Hamiltonian (\ref{eq:Hsemiinf}) is solved with the NRG
for the parameters $\varepsilon_{\rm f} = -U/2 = 10^{-3}$,
$\Lambda = 2.5$ and different values for $r$ and $\Delta_0$.
At each iteration step we keep $\approx 500$ states which
is sufficient for the calculation of thermodynamic properties.

We first want to discuss the phase-diagram of Fig.\ 1
where we have plotted the critical value $\Delta_{\rm c}$ versus $r$.
For any $\Delta_0 > \Delta_{\rm c}$ the system flows to a strong-coupling
fixed point (SC) similar to the fixed point in the standard case
\cite{Kri80}. The energy spectrum at this fixed point
can be explained by removing the first conduction electron site 
from the chain due to its strong coupling to the impurity.
The remaining chain, however, has a different structure as compared
to the $r\!=\!0$-case. Therefore this SC fixed point has not the Fermi liquid
properties of the standard single-impurity Anderson model (see below).
For $\Delta_0 < \Delta_{\rm c}$ the system always flows to the local-moment
fixed point (LM) with the impurity effectively decoupled from the
conduction band. Again, the resulting energy levels are in agreement
with those of the free conduction electron chain.

For both the Kondo model and the Anderson model $\Delta_{\rm c} (r)$
diverges at $r\!=\!\frac{1}{2}$ 
and we find for the Anderson model a logarithmic divergence
\begin{equation}
   \Delta_{\rm c,A} (r) \propto -\ln \left( \frac{1}{2} -r \right)
   .
\end{equation}
However, the behaviour of $\Delta_{\rm c} (r)$
for $0\!<\!r\!<\!\frac{1}{2}$ is quite different for both models.
Ingersent finds an extended linear region $\Delta_{\rm c} (r)\propto r$
which is approximately valid up to values of $r=0.4$
(see inset of Fig.\ 1).
In our case, $\Delta_{\rm c} (r)$ also starts linearly and is
in agreement with the result for the Kondo model up to
$r\approx 0.02$, but increases far more rapidly for larger $r$.

The difference to \cite{Ing96} is mainly due to the fact that for the
parameters used here, the f-level lies within the pseudogap density of
states. Under the assumption that the relevant coupling $\Delta^\prime$ for
this problem is (approximately) the value $\Delta(\omega=\varepsilon_{\rm
f})$ we have the exponential dependence
\begin{equation}
     \Delta^\prime(r) \approx \Delta_0 |\varepsilon_{\rm f}|^r =
     \Delta_0 e^{r\ln |\varepsilon_{\rm f}|} \label{eq:Delta^prime}.
\end{equation}
As $\ln |\varepsilon_{\rm f}|$ has a large negative value, $\Delta^\prime(r)$
is strongly supressed for increasing $r$ so that a much larger $\Delta_0$
is needed to reach the strong coupling fixed point. This increase of the
parameter regime in which local moment formation is observed has also been
found by Gonzalez-Buxton and Ingersent \cite{Gon96} who applied a poor man's
scaling approach to the Anderson version of the pseudogap problem.
To show that  eq.\ (\ref{eq:Delta^prime}) basically explains the difference
between the Kondo model and the Anderson model, we have plotted in the
inset of Fig.\ 1 both $\Delta_{\rm c,K}(r)$ for the Kondo model and
$\Delta_{\rm c,A}^\prime (r) =\Delta_{\rm c,A}(r) \cdot \exp (-7.9 \cdot r) $.
(the value 7.9 was chosen in order to fit $\Delta_{\rm c,A}^\prime (r)$
to $\Delta_{\rm c,K}(r)$)
The linear region of $\Delta_{\rm c,A}^\prime (r)$ now extends to
$r\approx 0.4$.

The remaining difference between $\Delta_{\rm c,K}(r)$ and 
$\Delta_{\rm c,A}(r)$ is due to the fact that 
the Kondo model and the Anderson model 
are related via the Schrieffer-Wolff transformation \cite{Sch66}
only in the limit $J\to 0$ (corresponding to $V^2/U \to 0$).
Therefore, the agreement of the results for both models is
only guaranteed for $\Delta \to 0$. Away 
from the line $\Delta = 0$, there is no exact mapping between the
Kondo version and the Anderson version.


The critical coupling $\Delta_{\rm c} (r)$ is determined as
follows. Fig.\ 2 shows the temperature dependence of the effective
magnetic moment for $U=0.001$, $\varepsilon_{\rm f}=-U/2$, $r=0.48$ 
and different values
of $\Delta$. In this graph, the LM fixed point (characterized by 
$\mu_{\rm res} \equiv \mu_{\rm eff}(T\to 0) = 1/4$) is reached within the 
given temperature range for
$\Delta = 0.01$ and $\Delta = 0.02$. The value $\mu_{\rm res} = r/8 = 0.06$
corresponding to the SC fixed point is clearly approached for  
$\Delta = 0.16$ while for $\Delta = 0.04$ this value should be reached at a
much lower temperature.
From Fig.\ 2 we determine $\Delta_{\rm c} (r=0.48)$ as
$\approx 0.03$ and repeat this procedure for different values of $r$.
Similar results for $\mu_{\rm eff} (T)$ have been obtained by
Ingersent and CY.

The value $\mu_{\rm res} = r/8$ at the SC fixed point can also be
derived directly from the semi-infinite chain form of the
free conduction electron chain at this fixed point.
One simply has to compare the effective magnetic moment for the system
with and without the first conduction electron site.

The temperature dependence of the specific heat coefficient
$\gamma(T)=C(T)/T$ in the SC regime
is shown in Fig.~3. The low temperature behaviour of  $\gamma(T)$
is described by a power law of the form
\begin{equation}
     \gamma(T) = c_1 T^{-r} + c_2 T^{-2r} . \label{eq:gamma2}
\end{equation}
Although eq.\ (\ref{eq:gamma2}) resembles an expansion in
$T^{-r}$, there cannot be any terms like $T^{-3r}$, $T^{-4r}$ etc.\
as the corresponding entropy would then diverge for $T\to 0 $
(for e.g.\ $r\!=\!0.4$).

The exponent $\alpha$ defined by $\gamma(T) \propto T^\alpha$
is shown in the inset of Fig.\ 3. In an intermediate temperature regime, 
$\alpha $ approaches the value $-r$ consistent  with the result
of CY.
However, for lower temperatures another term with the exponent
$\alpha = -2r$ is dominating.
This term is strongly suppressed in the intermediate regime
due to $c_2 \!<\!<\! c_1$.
This crossover from the $T^{-r}$ to the $T^{-2r}$
behaviour is {\it not} due to a crossover 
to a new low temperature fixed point.
For the spin susceptibility we confirm the result given by CY:
\begin{equation}
\chi(T)=  \frac{r}{8} 
 T^{-1} + c_1^\prime T^{-r} + c_2^\prime T^{-2r}.
\end{equation}

In the LM regime we find
\begin{equation}
     \gamma(T) = c_3 T^{r-1}  .
\end{equation}
and 
\begin{equation}
     \chi(T) = \frac{1}{4}T^{-1} + c_3^\prime T^{r-1}  .
\end{equation}

\section{Results for the spectral function}
The impurity spectral function
\begin{eqnarray}
   A(\omega) &=& \frac{1}{Z}\sum_{nm}  
           \bigg\vert   \Big< n \Big\vert f^\dagger_{-1\sigma}
                                         \Big\vert m \Big>
                   \bigg\vert^2
                   \delta \big( \omega -(E_{n} -E_{m}) \big) 
         \nonumber \\
         & & \hspace{1cm} \times \left( e^{-\beta E_m} + e^{-\beta E_n} \right),
    \label{eq:Ageneral}
\end{eqnarray}
(with the partition function $Z\!=\!\sum_m \exp(-\beta E_m)$),
has not yet been calculated in the previous papers on the pseudogap
problem.
We assume that the groundstate energy $E_g$ is set to zero and
concentrate on the zero-temperature limit, in which the 
spectral function takes the form
\begin{eqnarray}
  A(\omega) &=& \frac{1}{Z} \Big\{  2
              \sum_{n_g m_g}  
         \bigg\vert \Big< n_g \Big\vert f^\dagger_{-1\sigma}
           \Big\vert m_g \Big> \bigg\vert^2  \delta(\omega)  
              \nonumber \\
       & & \hspace{1cm} + \sum_{n_g m_e}  
         \bigg\vert \Big< n_g \Big\vert f^\dagger_{-1\sigma}
           \Big\vert m_e \Big> \bigg\vert^2  \delta(\omega+E_{m_e}) 
              \nonumber \\
       & & \hspace{1cm} + \sum_{n_e m_g}  
         \bigg\vert \Big< n_e \Big\vert f^\dagger_{-1\sigma}
           \Big\vert m_g \Big> \bigg\vert^2  \delta(\omega-E_{n_e}) \Big\}.
    \label{eq:Azero}
\end{eqnarray}
Here, the partition function $Z$ equals 
the total degeneracy of the groundstate. The $n_g,m_g$ label all states
with energy $E\!=\!E_g\!=\!0$ and the $n_e,m_e$ label the excited states.
The first term in eq.\ (\ref{eq:Azero}) would correspond to a transition
between different states with $E\!=\!0$, 
but 
such a term (resulting in a $\delta$-function at the Fermi level) is
not present in the NRG results.
There is one state with excitation energy $E_{\rm ex} \!\to\! 0$ for
$N\to\infty$ but its matrix element with the ground state
vanishes as $N\to\infty$. 

In order to obtain the full frequency dependence of the spectral function
within the NRG, it is necessary to combine the information of all iteration
steps as in each iteration the results are only given for 
a certain frequency range (see also \cite{Cos94,Sak89}).

In Fig.\ 4 we show results for the spectral function
for $r\!=\!0.25$, $\Delta > \Delta_{\rm c}$ (solid line, SC regime),
$r\!=\!0.25$, $\Delta < \Delta_{\rm c}$ (dotted line, LM regime) 
and $r\!=\!0.75$ (dashed line, LM regime). For these calculations we used
$\Lambda = 2$ and kept $\approx800$ states at each iteration.

We find that  $A(\omega)$ diverges as $|\omega|^{-r}$ for $\omega\to 0$
for any set of parameters which lies in the SC regime.
Note that this result suggests the conventional behaviour
$A(\omega)\sim 1/(\pi\Delta(\omega))$ as $\omega\to0$ for the SC case. Together
with the result $\gamma(T)\sim T^{-r}$ (neglecting the second term in (33) for
the moment) one could be tempted to interpret these results within a standard
Fermi liquid approach. Let us however emphasize that
in spite of these results the system is not a Fermi liquid for any $r>0$.

This observation becomes more evident in the LM regime, where the behaviour of
the spectral function is qualitatively different. Namely, in contrast to
the SC case we find that 
the spectral function vanishes as $A(\omega)\propto |\omega|^{r}$ here.
In addition, no qualitative difference, apart from the exponent, can be observed between
the cases $r\!>\!0.5$ and $r\!<\!0.5$.

\section{summary}
To summarize, we have studied the problem of an Anderson impurity in a
pseudo-gap Fermi system at particle-hole symmetry using a generalization of the
numerical renormalization group method \cite{Wil75,Kri80}.
We find a behaviour similar to that of the corresponding Kondo model
investigated by CY \cite{Che95} and Ingersent \cite{Ing96}.
However, the critical line $\Delta_{\rm c}$ separating the strong-coupling
and local-moment regimes of these two models shows a quite different form
between $r\!=\!0$ (where $\Delta_{\rm c}$ starts linearly) and
$r\!=\!1/2$ (where $\Delta_{\rm c}$ diverges). This difference
is mainly due to the fact that we have chosen the f-level 
to lie within the pseudogap.
In both strong-coupling and local-moment regime the thermodynamic
quantities specific heat and spin-susceptibility show powerlaw behaviour.

We also presented the first calculations of the impurity spectral function for
this model. We find $A(\omega)\propto |\omega|^{-r}$ in the  strong-coupling
regime and  $A(\omega)\propto |\omega|^r$ in the local-moment regime.
We do not find any indication that the local moment fixed points for
$r\!<\!1/2$ and $r\!>\!1/2$ are different.

As shown by Ingersent for the Kondo model, the critical values
$\Delta_{\rm c}$ take finite values as soon as particle-hole symmetry is
violated. It is of course interesting to see, whether this reduction of
the critical coupling is the same for the Anderson model (work on this
problem is in progress).

Another interesting question is the relevance of the model studied here in the
context of the dynamical mean field theory (for recent reviews see
\cite{Pru95,Geo96}). The effective single impurity
Anderson model appearing in the dynamical mean field 
theory is coupled to a (self-consistently
determined) effective medium. There is a possibility that the density of
states corresponding to this effective medium develops a pseudo-gap structure
under certain conditions (e.g.\ near the metal-insulator transition).
Also, the density of states of the infinite dimensional generalization
of the honeycomb ($d\!=2$) and diamond ($d\!=\!3$) lattices is
proportional to $|\omega|$ near the Fermi level. 

We wish to thank J.\ Keller and G.\ M.\ Zhang
for a number of stimulating discussions. One of us (R.B.) was supported
by a grant from the Deutsche Forschungsgemeinschaft, grant No.\ Bu965-1/1.

\appendix
\section{other discretization schemes}
In this appendix, we want to show that the discretization used by CY leads
to the same hopping matrix elements $t_n$ apart from a
redefinition of the discretization parameter
$\Lambda$.
This equivalence, however, is restricted to the special form of
$\Delta(\omega)$ and is not valid in general.

For $\Delta(\omega)=\Delta_0 |\omega|^r$ eq.\ (\ref{eq:pos2}) leads to
\begin{eqnarray}
   g(\varepsilon) &=& \varepsilon^{\frac{1}{r+1}} ,\\
   h^2 &=& \frac{\Delta_0}{\pi}\frac{1}{r+1}  \label{eq:hyb_app}.
\end{eqnarray}
The discrete energies $\xi_n^\pm$ take the form
\begin{eqnarray}
   \xi_n^+ &=& -\xi_n^- = \frac{1}{d_n} \int^+ {\rm d} \varepsilon g(\varepsilon)
   \nonumber \\
   &=& \frac{r+1}{r+2} \frac{1-\Lambda^{-\frac{r+2}{r+1}}}{
               1-\Lambda^{-1}} \Lambda^{-\frac{n}{r+1}}
   \nonumber \\
   &=& \frac{r+1}{r+2} \frac{1-\bar{\Lambda}^{-(r+2)}}{
               1-\bar{\Lambda}^{-(r+1)}} \bar{\Lambda}^{-n} \label{eq:xi_app} .
\end{eqnarray}
In the last equation we have defined $\bar{\Lambda} = 
\Lambda^{\frac{1}{r+1}}$.

The hybridization in eq. (\ref{eq:hyb_app}) is independent
of frequency,
therefore the result for $\left(\gamma_n^\pm\right)^2$ is the
same as for a constant $\Delta(\omega)=\Delta_0/(r+1)$
\begin{equation}
    \left(  \gamma_n^\pm \right)^2 =
     \frac{\Delta_0 }{r+1} \Lambda^{-n}
       \left(  1 - \Lambda^{-1} \right) . \label{gamma_appI}
\end{equation}   
With $\Lambda=\bar{\Lambda}^{(r+1)}$ eq.\ (\ref{gamma_appI}) gives
\begin{equation}
\left(  \gamma_n^\pm \right)^2 =  \frac{\Delta_0 }{r+1} 
  \bar{\Lambda}^{-n(r+1)}
       \left(  1 - \bar{\Lambda}^{-(r+1)} \right)  .
     \label{eq:gamma_app}
\end{equation}   
Eqs.\ (\ref{eq:xi_app}) and (\ref{eq:gamma_app}) correspond to 
Eqs.\ (\ref{eq:xi}) and (\ref{eq:gamma})  
with $\Lambda$ being replaced by $\bar{\Lambda}$. Obviously,
also the resulting matrix elements $t_n$ will have exactly the same
form as in eqs.\ (\ref{eq:tneven}) and (\ref{eq:tnodd}) and in terms
of $\Lambda$ we find in the limit of large $N$
\begin{equation}
      t_n \stackrel{n\to\infty}{\longrightarrow}
        \frac{r+1}{r+2} \,
        \frac{1-\Lambda^{-\frac{r+2}{r+1}}}{1-\Lambda^{-1}}
        \Lambda^{-\frac{n}{2(r+1)}} \left\{
           \begin{array}{lcl}
           1 &:& n\ {\rm even}\\
           \Lambda^{-\frac{r}{2(r+1)}} &:& n\ {\rm odd}
             \end{array}
            \right.    .  \label{eq:tnredCY}
\end{equation}   
For the ratio of $t_n/t_{n-1}$ we then find 
\begin{equation}
  \frac{t_n}{t_{n-1}} \stackrel{n\to\infty}{\longrightarrow}
   \left\{  \begin{array}{lcl} 
           \Lambda^{\frac{r-1}{2(r+1)}} &:& n\ {\rm even} \\
             \Lambda^{-1/2} &:& n\ {\rm odd}
          \end{array} \right. ,
\end{equation}   
corresponding to the result given in CY.

\begin{figure}[htb]
\unitlength1in
{\begin{picture}(3.8,3.8)
\epsfxsize=3.4in
\put(0.0,0){\epsffile{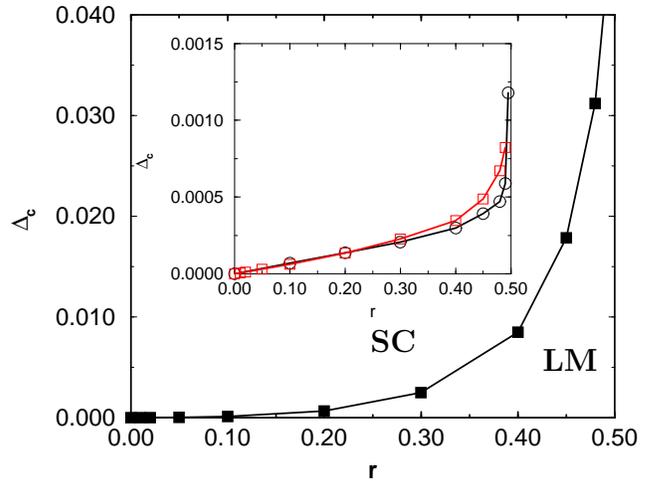}}
\epsfxsize=1.7in
\put(1.8,0.8){\large\bf SC}
\put(2.7,0.7){\large\bf LM}
\end{picture}
}
\caption{$r$-dependence of the critical coupling $\Delta_{\rm c}$
  which seperates the strong-coupling regime ($\Delta > \Delta_{\rm c}$)
  from the local-moment regime ($\Delta > \Delta_{\rm c}$). 
  The filled squares show the result for $\Delta_{\rm c,A}(r)$.
  In the inset we compare the scaled critical coupling 
  $\Delta_{\rm c,A}^\prime(r)$ (open squares) with the result obtained
  by Ingersent for the Kondo version of the Hamiltonian (3) (circles).
}
\label{fig:Delta_c}
\end{figure}

\begin{figure}[htb]
\epsfxsize=3.4in
\epsffile{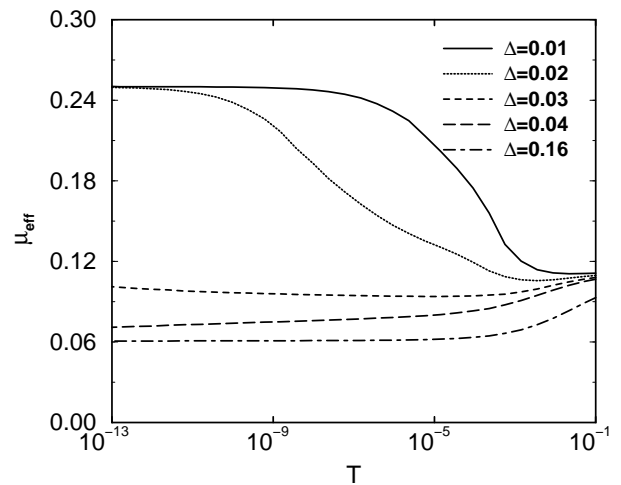}
\caption{Temperature dependence of the effective magnetic moment
  for $U=0.001$, $\varepsilon_{\rm f}=-U/2$, $r=0.48$ and various values
  of $\Delta$. For $\Delta < 0.03$, the system flows to the 
  local-moment fixed point with the corresponding effective magnetic moment 
  $\mu_0 = 1/4$. For $\Delta > 0.03$ the magnetic moment 
  is partially screened and approaches the residual value $\mu_{\rm res} 
  = r/8 = 0.06$ from above.}
\end{figure}

\begin{figure}[htb]
\epsfxsize=3.4in
\epsffile{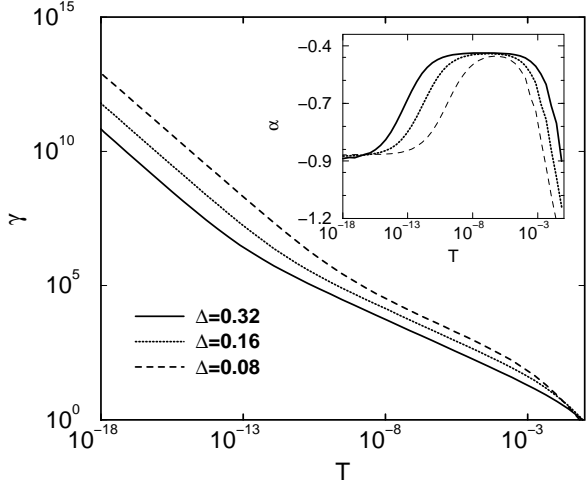}
\caption{Temperature dependence of the specific heat coefficient
  $\gamma(T) = C(T)/T$ for $U=0.001$, $\varepsilon_{\rm f}=-U/2$, 
  $r=0.48$ and various values
  of $\Delta$. The inset shows the temperature dependence of the 
  exponent $\alpha(T)$ defined by
  $\gamma(T) \propto T^\alpha$. In an intermediate regime, $\alpha$ approaches $-r$ as
  expected from eq.\ (1) but for lower temperatures a stronger divergent
  term with $\alpha\approx -0.93 \approx -2r$ dominates. This behaviour is not due
  to the crossover to a new fixed point.}
\end{figure}

\begin{figure}[htb]
\epsfxsize=3.4in
\epsffile{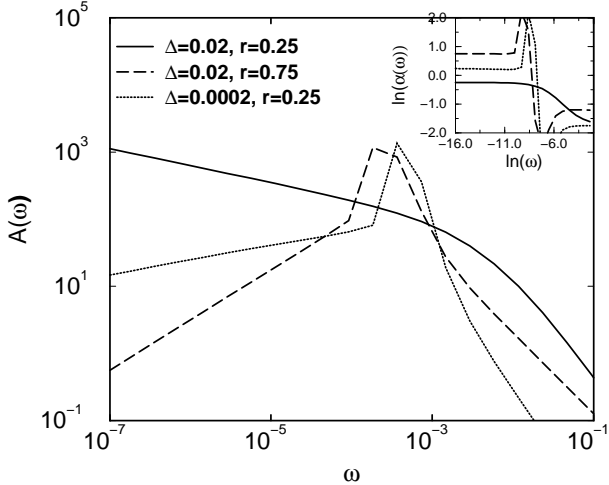}
\caption{Impurity spectral function for $U=0.001$,
   $\varepsilon_{\rm f}=-U/2$, and different values of $r$ and $\Delta$.
  for $r\!=\!0.25$ and $\Delta=0.02$ (solid line, SC-regime), the
  spectral function diverges as $A(\omega)  \propto|\omega|^{-r}$
  for $|\omega|\to 0$. In the LM-regime (for both 
  $r\!=\!0.25$ and $\Delta=0.0002$ (dotted line) and 
  $r\!=\!0.75$ and $\Delta=0.02$ (dashed line)) the  spectral function
  vanishes as $|\omega|^{r}$.
  The inset shows the coefficient $\alpha(\omega)$ defined by 
  $A(\omega)  \propto|\omega|^{\alpha(\omega)}$.}
\end{figure}


\begin{thebibliography}{ABC99a}
\bibitem{comment} The restriction to a hybridization depending on $k$ only
through
$\varepsilon_k$ is not necessary but convenient for pedagogical reasons.
\end{thebibliography}
\end{document}